\documentclass[
reprint,
superscriptaddress,
amsmath,amssymb,
aps,
showpacs
]{revtex4-1}

\usepackage{graphicx}
\usepackage[colorlinks=true,citecolor=blue,linkcolor=blue,urlcolor=blue,linktocpage]{hyperref}

\begin{document}

\title{Chirp mitigation of plasma-accelerated beams using a modulated plasma density}

\author{R. Brinkmann}
\affiliation{Deutsches Elektronen-Synchrotron DESY, Notkestr.\ 85, 22607 Hamburg, Germany }
\author{N. Delbos}
\affiliation{Center for Free-Electron Laser Science $\&$ Department of Physics University of Hamburg, Luruper Chaussee 149, 22761 Hamburg, Germany}
\author{I. Dornmair}
\affiliation{Center for Free-Electron Laser Science $\&$ Department of Physics University of Hamburg, Luruper Chaussee 149, 22761 Hamburg, Germany}
\author{R. Assmann}
\affiliation{Deutsches Elektronen-Synchrotron DESY, Notkestr.\ 85, 22607 Hamburg, Germany }
\author{C. Behrens}
\affiliation{Deutsches Elektronen-Synchrotron DESY, Notkestr.\ 85, 22607 Hamburg, Germany }
\author{K. Floettmann}
\affiliation{Deutsches Elektronen-Synchrotron DESY, Notkestr.\ 85, 22607 Hamburg, Germany }
\author{J. Grebenyuk}
\affiliation{Deutsches Elektronen-Synchrotron DESY, Notkestr.\ 85, 22607 Hamburg, Germany }
\author{M. Gross}
\affiliation{Deutsches Elektronen-Synchrotron DESY, Platanenallee 6, 15738 Zeuthen, Germany }
\author{S. Jalas}
\affiliation{Center for Free-Electron Laser Science $\&$ Department of Physics University of Hamburg, Luruper Chaussee 149, 22761 Hamburg, Germany}
\author{M. Kirchen}
\affiliation{Center for Free-Electron Laser Science $\&$ Department of Physics University of Hamburg, Luruper Chaussee 149, 22761 Hamburg, Germany}
\author{T. Mehrling}
\affiliation{Deutsches Elektronen-Synchrotron DESY, Notkestr.\ 85, 22607 Hamburg, Germany }
\author{A. Martinez de la Ossa}
\affiliation{Deutsches Elektronen-Synchrotron DESY, Notkestr.\ 85, 22607 Hamburg, Germany }
\author{J. Osterhoff}
\affiliation{Deutsches Elektronen-Synchrotron DESY, Notkestr.\ 85, 22607 Hamburg, Germany }
\author{B. Schmidt}
\affiliation{Deutsches Elektronen-Synchrotron DESY, Notkestr.\ 85, 22607 Hamburg, Germany }
\author{V. Wacker}
\affiliation{Deutsches Elektronen-Synchrotron DESY, Notkestr.\ 85, 22607 Hamburg, Germany }
\author{A. R. Maier}
\email{andreas.maier@desy.de}
\affiliation{Center for Free-Electron Laser Science $\&$ Department of Physics University of Hamburg, Luruper Chaussee 149, 22761 Hamburg, Germany}

\date{\today}

\begin{abstract}
Plasma-based accelerators offer the possibility to drive future compact light sources and high- energy physics applications. Achieving good beam quality, especially a small beam energy spread, is still one of the major challenges. For stable transport, the beam is located in the focusing region of the wakefield which covers only the slope of the accelerating field. This, however, imprints a longitudinal energy correlation (chirp) along the bunch. Here, we propose an alternating focusing scheme in the plasma to mitigate the development of this chirp and thus maintain a small energy spread.
\end{abstract}

\maketitle

Using the extreme field gradients supported by a plasma-cavity \cite{cite:Veksler1956}, plasma-based accelerators \cite{cite:Esarey2009} promise very compact sources of ultra-relativistic electron beams for a large variety of applications. Yet, especially the beam energy spread, which is in plasma experiments typically on the percent level \cite{cite:rechatin2009}, causes emittance growth during beam transport, and hence, renders its applicability to novel FEL schemes \cite{cite:maier2012,cite:huang2012} or high-energy physics applications very difficult. Controlling the beam energy spread is thus one of the major challenges in the field of plasma acceleration.

External injection of a well-characterized and tuned electron beam into a plasma acceleration stage is a promising concept towards high-quality beams. Decoupling the generation of the beam from the acceleration allows to independently control and optimize the dynamics of each process. Moreover, it is an integral part of any staged acceleration scheme aiming towards highest beam energies \cite{cite:Schroeder2010}. The coupling of an external electron beam into and out of a plasma accelerator stage, while preserving the beam quality, has been recently discussed \cite{cite:Assmann1998, cite:Mehrling2012, cite:Dornmair2015}.

During the acceleration inside the plasma the beam is typically located at the slope of the accelerating field (referred to as off-crest acceleration), such that it is simultaneously accelerated and focused by the plasma fields. However, this choice of accelerating phase also imprints a longitudinal energy correlation (chirp) onto the bunch -- an intrinsic feature of virtually all plasma-acceleration schemes, and a major source of the undesired energy spread growth.

In this Letter, we propose a novel scheme based on alternating focusing, which mitigates the energy chirp accumulation in the plasma.  By modulating the plasma density, the bunch periodically experiences accelerating fields with opposite slope, which effectively suppresses the chirp evolution. Although the bunch is thereby alternatingly focused and defocused, we show that stable beam transport can be achieved and the beam quality can be preserved, similar to the concept proposed by Courant and Snyder \cite{cite:Courant1952}. 

The paper is structured as follows: First, we discuss the longitudinal and transverse fields in a plasma with periodically modulated density, and derive the periodic beta function for stable transport from a matrix formalism. We then demonstrate the chirp mitigation of our alternating focusing (AF) scheme using Particle-In-Cell (PIC) simulations for a laser- and a beam-driven plasma wakefield, and compare it to a reference case of constant plasma density.

In the following, we assume a linear plasma wakefield generated by a Gaussian driver described by $f(r,\zeta) = f_0\exp{\left(-r^2/2\sigma_r^2\right)}\exp{\left(-\zeta^2/2\sigma_z^2\right)}$. Our discussion applies to both laser driven wakefields, where $f_0=a_0^2/4$ for a linearly polarized laser pulse, and particle beam driven wakefields, with $f_0=n_b/n_e$ for an electron beam. Here, $n_b$ is the density of the driver beam, while $n_e$ denotes the plasma electron density. The plasma wave number is $k_p=\omega_p/c$, with $\omega_p=(n_ee^2/\epsilon_0 m_e)^{1/2}$, and $\zeta = z - ct$ the distance behind the driver. The peak normalized laser vector potential is $a_0 = eA/m_ec^2$. In the linear wakefield regime, $a_0^2 \ll 1$ for a laser driver, and $n_b/n_e \ll 1$ for an electron beam driver. In an external injection scheme, the electron bunch may be positioned at an arbitrary phase $\Psi=k_p\zeta$ of the plasma wave. The wakefields far behind the driver are \cite{cite:Esarey2009, cite:Gorbunov1987}
\begin{eqnarray}
E_z(r,\zeta) &=& E_{z,0} \exp{\left(-\frac{r^2}{2\sigma_r^2}\right)} \cos{(k_p\zeta)}, \label{eq:EzEr}\\
E_r(r,\zeta)-cB_\theta(r,\zeta) &=& -E_{z,0}\frac{r}{k_p\sigma_r^2} \exp{\left(-\frac{r^2}{2\sigma_r^2}\right)} \sin{(k_p\zeta)}, \nonumber
\end{eqnarray}
with $E_{z,0}$ the amplitude of the longitudinal field component. The focusing strength $K$ acting on a relativistic electron bunch in the vicinity of the axis, with $\gamma_L$ the relativistic Lorentz factor of the accelerated electron bunch, is
\begin{equation}
K(\zeta) = \frac{e}{\gamma_L m_e c^2} \left.\partial_r \left( E_r-cB_\theta\right)(r,\zeta)\right|_{r=0}.
\label{eq:K}
\end{equation}

It would be desirable to position the accelerated electron bunch on-crest, i.e.\ at the maximum of the accelerating field at $\Psi=-\pi$, which avoids, to first order, a gradient of the accelerating field along the bunch. Yet, for plasma accelerators, the bunch tail would erode due to the defocusing fields for $\Psi<-\pi$. For off-crest acceleration, within the interval of $-\pi < \Psi < -\pi/2$, the transverse fields focus the whole electron beam, while it is accelerated by the longitudinal field $E_z(\zeta)$. This allows stable transport of the beam, albeit at the cost of a correlated energy spread, caused by the gradient in $E_z$.

\begin{figure}
\includegraphics[width=1.0\columnwidth]{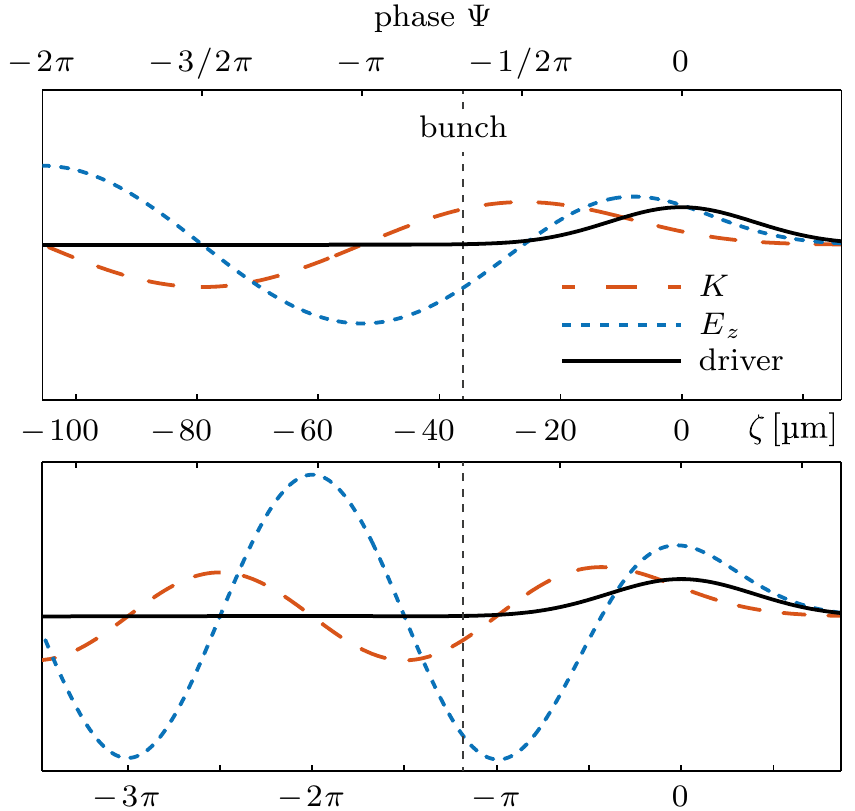}
\caption{On-axis accelerating field $E_z$ [arb. u.] and focusing strength $K$ [arb. u.] in a plasma cavity, following eqs.\ \eqref{eq:EzEr} and \eqref{eq:K}, for a density $n_\text{min}=1\times10^{17}\text{cm}^{-3}$ (top) and $n_\text{max}=3\,n_\text{min}$ (bottom). This corresponds to a plasma wavelength of $106\,\mathrm{{\mu}m}$ and $61\,\mathrm{{\mu}m}$, respectively.}
\label{fig:n1n2}
\end{figure}

The effect can be mitigated with smaller ratios of $\sigma_z/\lambda_p$, i.e. using shorter bunch lengths $\sigma_z$, or lower plasma densities. Yet, ever shorter electron bunches are difficult to generate in a conventional RF-based accelerator, and also they support much smaller beam charge. In addition, lower plasma densities require significantly higher laser pulse energies to resonantly drive the wake, limit the bunch density as beamloading effects become more severe, and decrease the accelerating gradient. Alternatively, it has been proposed to utilize the beam loading field of the electron beam in the plasma to flatten $E_z$ along the bunch \cite{cite:Katsouleas1987}. However, this requires very precise control over the injected current profile to exactly cancel the gradient in $E_z$ along the bunch, and has, so far, not been demonstrated with the desired controllability and beam quality.

As a novel approach to this problem, we illustrate our concept of chirp mitigation via alternating focusing in figure \ref{fig:n1n2}. The top panel shows the fields in the plasma cavity, eqs.\ \eqref{eq:EzEr} and \eqref{eq:K}, for a density $n_\text{min}=1\times10^{17}\text{cm}^{-3}$, where the bunch is at an accelerating and focusing phase. For a fixed delay $\zeta/c$, but higher density $n_\text{max}=3\,n_\text{min}$, the bunch is defocused, but accelerated by a field of opposite slope and higher amplitude. Shifting the bunch between both phases with a periodically modulated plasma density then allows to mitigate chirp evolution.

Successful implementation of the concept requires stable beam transport through the periodically focusing and defocusing regions of the plasma. We now derive a condition for stable beam transport. In general, an electron under the influence of a constant focusing force $K$ is propagated along $z$ by the transport matrix $M$,
\begin{equation}
\begin{pmatrix}    r\\    r' \end{pmatrix} = M \cdot \begin{pmatrix}    r_0\\    r'_0 \end{pmatrix},
\quad \text{with} \quad
M = \begin{pmatrix}    C \ S\\    C'\ S' \end{pmatrix}.
\nonumber
\end{equation}
Here, $C=\cos(\sqrt{K}z)$ and $S= K^{-1/2} \sin(\sqrt{K}z)$. To describe the transport over one period $\lambda_\text{mod}$ of the modulated plasma density, we approximate the focusing profile $K=K(z)$ by a sequence of $i=1, ..., n$ stepwise constant $K_i$. The transport matrix over $\lambda_\text{mod}$ is then $T= M_n \cdot ... \cdot M_1$. Stable transport requires $|\mathrm{Tr}(T)| < 2$ \cite{cite:courant1958}. To describe the evolution of the whole electron beam, we use the Courant-Snyder parameters $\alpha= -\langle xx' \rangle/\epsilon$, $\beta = \langle x^2 \rangle/\epsilon$ and $\gamma = (1+\alpha^2)/\beta$, with $\epsilon$ the transverse emittance. They are related to the transport matrix
\begin{equation}
T =\begin{pmatrix}    a \ b\\    c \ d \end{pmatrix} = \mathbb{I} \cos(\mu) + J \sin(\mu),\; \text{with} \; J = \begin{pmatrix}    \alpha \quad \beta,\\    -\gamma \ -\alpha \end{pmatrix}
\nonumber
\end{equation} 
with $\mathbb{I}$ being the identity matrix, $\mu$ the phase advance, $2\cos(\mu) = \mathrm{Tr}(T)$, and $\sin(\mu) = \sqrt{1 - \cos^2(\mu)}$. From this, we can derive the beta function at the beginning of the periodic structure, in our case the beginning of the modulated plasma, that is needed for stable transport 
\begin{equation}
\beta(\zeta) = \frac{2b}{\sqrt{4-(a+d)^2}}.
\label{eq:beta}
\end{equation}

Since $K$ depends on $\zeta$, the distance between bunch and driver, also the transport matrix $T$ and consequently $\beta$ depend on $\zeta$. Real values of $\beta(\zeta)$ correspond to stable beam transport. The net accelerating field at $\zeta$ is $\bar{E}_z(\zeta) = \lambda_\text{mod}^{-1} \int_0^{\lambda_\text{mod}} E_z(r=0,\zeta,z)\,\mathrm{d}z$. We choose the position $\zeta$ of the electron bunch behind the driver such, that (i) we obtain stable transport, eq.\ \eqref{eq:beta}, and (ii) the bunch is accelerated on-crest of the averaged accelerating field. There, at the minimum of $\bar{E}_z$, we maximize the energy gain and mitigate the chirp evolution.

To demonstrate the alternating focusing scheme, we assume in the following a driver laser pulse of Gaussian shape, with $a_0=0.8$, $\sigma_z=9.7\,\mu$m, and a spot size of $w_0 = 2\sigma_r = 50\,\mu$m. We further assume a plasma density modulated between $n_\text{max} = 3\times10^{17}\,\text{cm}^{-3}$ and $n_\text{min} = 1\times10^{17}\,\text{cm}^{-3}$, and a modulation profile,
\begin{equation}
k_p(z) = k_{p,\text{min}} + (k_{p,\text{max}} - k_{p,\text{min}}) \cdot |\sin(\pi z/\lambda_\text{mod})|^b,
\nonumber
\end{equation}
with $k_{p}\propto\sqrt{n_e}$ and modulation period $\lambda_\text{mod}=1$\,mm. The parameter $b$ describes the steepness and width of the density modulation, with $b=4$ in our example. Note that the high density region is shorter than the low density region: The varying plasma density not only changes the phase of the bunch within the wakefield, but also sign and steepness of the accelerating field gradient. Therefore, the compensation of the chirp requires a larger fraction of $\lambda_\text{mod}$ than its buildup.

\begin{figure}
\includegraphics[width=1.0\columnwidth]{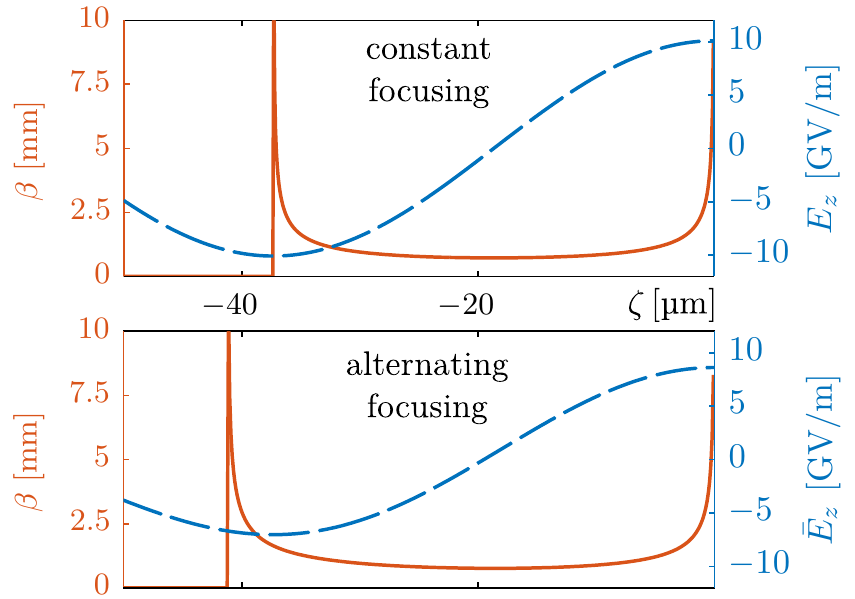}
\caption{Matched beta function (solid red) and net accelerating field (dashed blue), for a constant plasma density (top) and a modulated plasma (bottom), for which we find a distance $\zeta\approx-38\,\mu$m, equivalent to on-crest acceleration and stable transport.}
\label{fig:matchedBeta}
\end{figure}

Using this density profile, we calculate $\beta(\zeta)$ and $\bar{E}_z(\zeta)$, shown in figure \ref{fig:matchedBeta} (bottom), and compare it with a reference case of constant density $n_e = 2\times 10^{17}\,\mathrm{cm^{-3}}$ (top). Both cases have a singularity in the beta function, which marks the transition from regions of unstable to stable transport. In the reference case (top), all delays with stable beta function have a gradient in the accelerating field and would thus imprint a chirp on the accelerated bunch. Contrarily, in the alternating focusing case (bottom), we find a delay with a minimum in $\bar{E}_z$ and a stable beta function. This delay is equivalent to on-crest acceleration in a conventional accelerator, and no chirp is imprinted on the bunch.

To validate our results, we perform 3D particle-in-cell (PIC) simulations using the code \textsc{Warp} \cite{cite:Friedman2014} in the Lorentz boosted frame. The simulation box volume is $122\,\mathrm{{\mu}m}\times200\,\mathrm{{\mu}m}\times200\,\mathrm{{\mu}m}$ with $4892\times100\times100$ cells and one particle per cell. The simulation is boosted by $\gamma_\text{boost} = 8$. We inject an external electron beam of 100\,MeV, with a bunch length of $\sigma_z=1\,\mu$m, and a normalized emittance of $\epsilon_n=0.5$\,mm.mrad. The transverse size of $\sigma_{x/y}=2\,\mu$m is matched to the alternating focusing structure, using the simulated focusing forces and eq.\ \eqref{eq:beta}. The bunch has a relative energy spread of $\sigma_\gamma/\gamma=0.1\,\%$, no longitudinal chirp, and a bunch charge of $1\,$pC to avoid beam loading effects. The distance between driver laser and witness bunch is chosen slightly larger than the analytical estimation with $\Delta\zeta = -40\,\mathrm{{\mu}m}$ to account for the relativistically elongated plasma wavelength and slippage effects. For reference, we perform a simulation with the same parameters but with a constant flat-top plasma profile of density $n_e=2\times 10^{17}\,\text{cm}^{-3}$, and here $\Delta\zeta = -39\,\mathrm{{\mu}m}$.

\begin{figure}
\includegraphics[width=1.0\columnwidth]{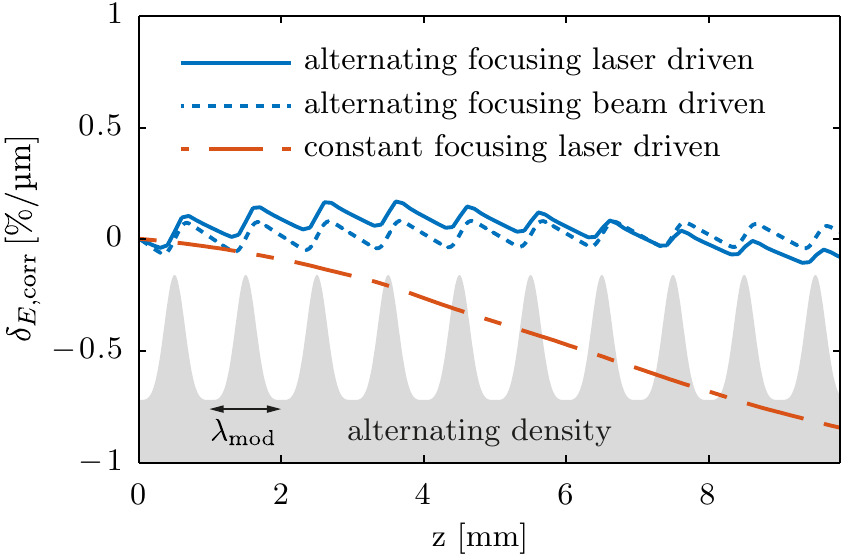}
\caption{Evolution of the correlated energy spread (chirp) $\delta_{E,\text{corr}}=\langle zE_\text{kin} \rangle / \langle z^2 \rangle \bar E_\text{kin}$. The chirp evolution for the constant focusing reference case is indicated in red (dashed-dotted), while the blue line (solid) shows the chirp evolution using the modulated plasma density. In contrast to the laser-driven case, there is no slippage between driver and witness electron bunch, in an electron beam driven wakefield (blue dotted), and we see no global variation in the correlated energy spread. The grey shaded area indicates the modulated plasma density profile.}
\label{fig:chirp}
\end{figure}

Figure \ref{fig:chirp} shows the evolution of the correlated energy spread (chirp) for both cases. In the reference case (red dashed) the bunch is in the focusing region, thus off-crest of the accelerating field, and develops a steadily increasing energy chirp. In contrast, using a modulated plasma density, the energy chirp is compensated after every modulation period $\lambda_\text{mod}$. As the electron bunch is faster than the laser in the plasma, it slowly slips through the minimum in $\bar{E}_z(\zeta)$, compare figure \ref{fig:matchedBeta} (bottom), which causes the small global variation of the chirp. This effect could, however, be compensated with a globally tapered plasma density \cite{cite:rittershofer2010}. Over one $\lambda_\text{mod}$ the chirp builds up faster than it decreases, which is a consequence of the aforementioned asymmetry in plasma density and accelerating field gradient. The beam emittance is conserved throughout the entire plasma.

\begin{figure}
\includegraphics[width=1.0\columnwidth]{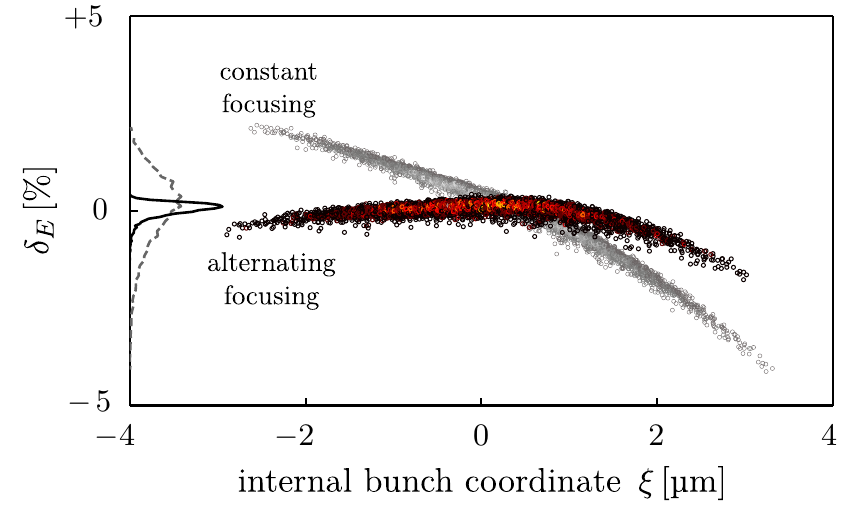}
\caption{Electron beam phase space after the propagation through the plasma. Here, $\delta_E = \Delta E_\text{kin}/\bar E_\text{kin}$. In the reference case (grey) the beam has accumulated a negative chirp, while the chirp is compensated using a modulated plasma density. For the chosen parameter set the mean beam energy is $\bar E_\text{kin}=157$\,MeV for the alternating focusing scheme, and $\bar E_\text{kin}=181\,$MeV for the reference case.}
\label{fig:PS}
\end{figure}

We further investigate the alternating focusing scheme for an electron beam driven wakefield with 3D simulations using the PIC code HiPACE \cite{cite:Mehrling2014} with a simulation box volume of $118\,\mathrm{{\mu}m}\times235\,\mathrm{{\mu}m}\times235\,\mathrm{{\mu}m}$ with $512\times320\times320$ cells. For this purpose, we assume a Gaussian driver with $\sigma_z= 9.7\,\mu$m,  $\sigma_{r}=25\,\mu$m, and $\epsilon_{n}=2.5\,$mm.mrad. The driver peak density of $n_b = 2.6\times10^{16}~\mathrm{cm}^{-3}$ corresponds to a charge of \mbox{400 pC}. For simplicity, we choose in this conceptual study a driver beam energy of $20\,\text{GeV}$, to prevent an evolution of the driver, with an energy spread of $0.1\,\%$.

The witness bunch of \mbox{0.01 pC} has a Gaussian shape with $\sigma_{x/y}=2\,\mu$m, $\sigma_z=1\,\mathrm{\mu m}$, and $\epsilon_n=0.5\, \text{mm.mrad}$. Its initial energy is $100\,\text{MeV}$ with an initially uncorrelated energy spread of $\sigma_\gamma/\gamma=0.1\,\%$.

We use the same modulated plasma profile as before and select the driver beam density such that the ratio $f_0 = n_b/n_e(\bar k_p)=a_0^2/4$ is the same as in the laser driven case, which makes both comparable. Analog to the laser driven case, we calculate $\beta(\zeta)$ and $\bar E_z(\zeta)$, and choose $\Delta\zeta = -45\,\mathrm{{\mu}m}$ such that the witness is in the minimum of $\bar E_z$.

As expected, our simulations show, that the energy chirp of the accelerated bunch is compensated, as in the laser-driven case, see figure \ref{fig:chirp}. Since there is no slippage between witness and driver, we find no modulation in the global chirp evolution, and conclude that our scheme works for both laser and beam driven scenarios.

In figure \ref{fig:PS}, we compare the witness beam phase space of the laser-driven AF case to the reference case at the end of the plasma. Using the alternating focusing scheme, the bunch has virtually no correlated energy spread, unlike the reference case (grey). The projected rms energy spread is reduced by a factor of 4 to $0.24\,\%$, compared to $0.96\,\%$ in the reference case.

As presented here, our concept is based on the linear wakefield regime. The approach, however, is general, and could also be applied to different plasma densities, driver properties, and to the non-linear regime. For the latter, further studies are necessary to determine limitations arising from the shortening of the defocusing region and the changed focusing forces.

In conclusion, we have proposed a novel alternating focusing scheme for laser- and beam-driven wakefield accelerators, based on a modulated plasma density profile. Our concept effectively suppresses the buildup of large energy chirps during acceleration, while conserving beam emittance, which is crucial for any staged acceleration scheme. Thus, it promises to overcome one of the major challenges in the field, the generation of electron beams of a quality that rivals and possibly outperforms those from traditional radio-frequency accelerators. By providing the freedom to shape the net accelerating field along the bunch, alternating focusing may be the key concept to unleash the potential of plasma-based accelerator technology for applications.

We gratefully acknowledge the computing time provided on the supercomputer JURECA under project HHH20. We acknowledge the use of the High-Performance Cluster (IT-HPC) at DESY. This work was funded in part by the Humboldt Professorship of B. Foster, the DAAD, and the Helmholtz Virtual Institute VH-VI-503.

\end{document}